\documentclass[conference]{IEEEtran}
\IEEEoverridecommandlockouts
\usepackage{cite}
\usepackage{amsmath,amssymb,amsfonts}
\usepackage{algorithmic}
\usepackage{graphicx}
\usepackage{textcomp}
\usepackage{xcolor}
\usepackage{listings}
\usepackage{graphicx}
\usepackage[final]{microtype}
\usepackage{url}
\def\BibTeX{{\rm B\kern-.05em{\sc i\kern-.025em b}\kern-.08em
    T\kern-.1667em\lower.7ex\hbox{E}\kern-.125emX}}
\begin{document}

\lstdefinestyle{mystyle}{
    backgroundcolor=\color{white},
    commentstyle=\color{gray},
    keywordstyle=\color{blue},
    numberstyle=\tiny\color{gray},
    stringstyle=\color{orange},
    basicstyle=\ttfamily\footnotesize,
    breaklines=true,
    captionpos=b,
    keepspaces=true,
    numbers=left,
    numbersep=8pt,
    xleftmargin=15pt,
    showspaces=false,
    showstringspaces=false,
    showtabs=false,
    tabsize=2
}

\lstset{style=mystyle}

\title{Consensus Is All You Need: Gossip-Based Reasoning Among Large Language Models\\}

\author{\IEEEauthorblockN{Saksham Arora}
\IEEEauthorblockA{
Sunnyvale, California \\
sakshamarora239@gmail.com \\
https://orcid.org/0009-0008-4566-7518}
}

\maketitle

\begin{abstract}
Large language models have advanced rapidly, but no single model excels in every area—each has its strengths and weaknesses. Instead of relying on one model alone, we take inspiration from gossip protocols in distributed systems, where information is exchanged with peers until they all come to an agreement. In this setup, models exchange answers and gradually work toward a shared solution. Each LLM acts as a node in a peer-to-peer network, sharing responses and thought processes to reach a collective decision. Our results show that this "gossip-based consensus" leads to robust, resilient, and accurate multi-agent AI reasoning. It helps overcome the weaknesses of individual models and brings out their collective strengths. This approach is similar to how humans build consensus, making AI seem more collaborative and trustworthy instead of just a black-box program.
\end{abstract}


\section{Introduction}
Large language models like GPT-4\cite{openai2023gpt4}, Claude\cite{anthropic2023claude}, Gemini\cite{gemini2024}, DeepSeek\cite{deepseek2024}, and Grok\cite{grok2024} have demonstrated impressive capabilities across a wide range of tasks. However, they differ in architecture, training data, biases, and response behavior, leading to inconsistent performance. Inspired by consensus methods in distributed systems, we use gossip protocols as a way for different models to interact and reach agreement. Here, every LLM acts like a peer in the network—producing answers, sharing them with others, and updating its view based on feedback. We try to use different pipelines to gather results—by using the gossip protocol, we can enable multiple different architectures for how these LLMs converge to an answer. The gossip protocol enables nodes to communicate with each other, exchanging local information and updating their views. We can also control how many rounds of communication occur before the system reaches a high-confidence consensus.

There are many ways we can make this algorithm work—by refining answers through prompt engineering and then performing the voting process. Though weighted voting can be applied, it often introduces bias, contradicting our goal of decentralizing inference authority. As such, our approach mirrors modern human consensus-building by avoiding centralized dominance.

For instance, with a large number of LLMs, we can form multiple subsets (e.g., three sets of three models each). Each set first reaches an internal consensus, then leaders from each group propose their answers to a final layer of consensus. This structure helps optimize for limited context windows, particularly for models with smaller memory capacity.

During consensus, we also pass not just final answers but the thought processes of each model—how they arrived at their decisions. Metadata about previous rounds (e.g., why an answer was selected) can also be shared in multi-round iterations. This metadata allows us to explore whether models develop preferences or implicit biases—whether they tend to favor certain peers more frequently. Findings show that this gossip framework can be really simple and also really complex, according to the application needs.

\section{Motivation and Background}

\subsection{Motivation}
\begin{itemize}
    \item \textbf{Decentralization:} Avoids dependence on a single model, utilizing all available models.
    \item \textbf{Redundancy and fault-tolerance:} Local mistakes can be corrected via peer feedback.
    \item \textbf{Scalability:} Easily accommodates new models without retraining or architecture changes.
    \item \textbf{Emergent agreement:} Consensus emerges from iterative interaction, mimicking social and biological systems \cite{kempe2003gossip,renesse1998gossip}.
    \item \textbf{Human-like collaboration:} Reflects how humans collectively reason and deliberate.
\end{itemize}

\subsection{Background}
Gossip protocols, modeled after how information spreads in social groups, where each node in the system represents a peer, and repeated rounds of information exchange quickly synchronize the network. This repeated, decentralized sharing gradually aligns the global state without requiring strict synchronization. Such techniques have powered some of the largest and most resilient systems—like DynamoDB and Cassandra—for maintaining eventual consistency \cite{lakshman2010cassandra}.

Beyond the technical motivations, it is important to acknowledge the human-facing implications of collaborative AI systems. As these artificial intelligence systems becomes more integrated into everyday life, it is critical that people do not perceive these systems as opaque or threatening. A consensus-based system inspired by human decision-making processes fosters trust and transparency. If models are trained to behave more like collaborative human agents—deliberating, debating, and arriving at consensus—the resulting system aligns better with human social intuition.

This approach opens the door to building AI systems that feel less like black boxes and more like thoughtful collaborators. In group decision-making contexts—whether in business, governance, or healthcare—humans rarely rely on a single authority. Instead, ideas are debated, rationales are exchanged, and decisions are shaped collectively. Mimicking this process allows models to become relatable and interpretable, giving users the sense that they are engaging with an ensemble of diverse perspectives rather than a singular, monolithic output.

Such systems can also help correct for individual model bias and present disagreements in a manner that promotes user confidence. Rather than being perceived as inflexible or authoritarian, these models can be seen as co-thinkers: part of a collective that values reflection, dialogue, and shared reasoning.

We extend this principle to AI: rather than propagating data states, our gossip system propagates belief states and thought processes, ultimately reaching consensus not on bytes—but on judgment.

\section{Algorithm Design}
We define three primary variants of the gossip consensus mechanism for multi-agent LLM collaboration.

\subsection{Simple Voting Algorithm (With Context)}
\textbf{Input:} A set of models $\{M_1, M_2, M_3\}$ and a question $Q$.

\begin{itemize}
    \item Each model generates an answer and corresponding thought process.
    \item During the consensus round, all models receive the responses from their peers (excluding their own).
    \item A majority vote determines the final answer.
\end{itemize}

\begin{figure}[ht]
    \centering
    \includegraphics[width=0.45\textwidth]{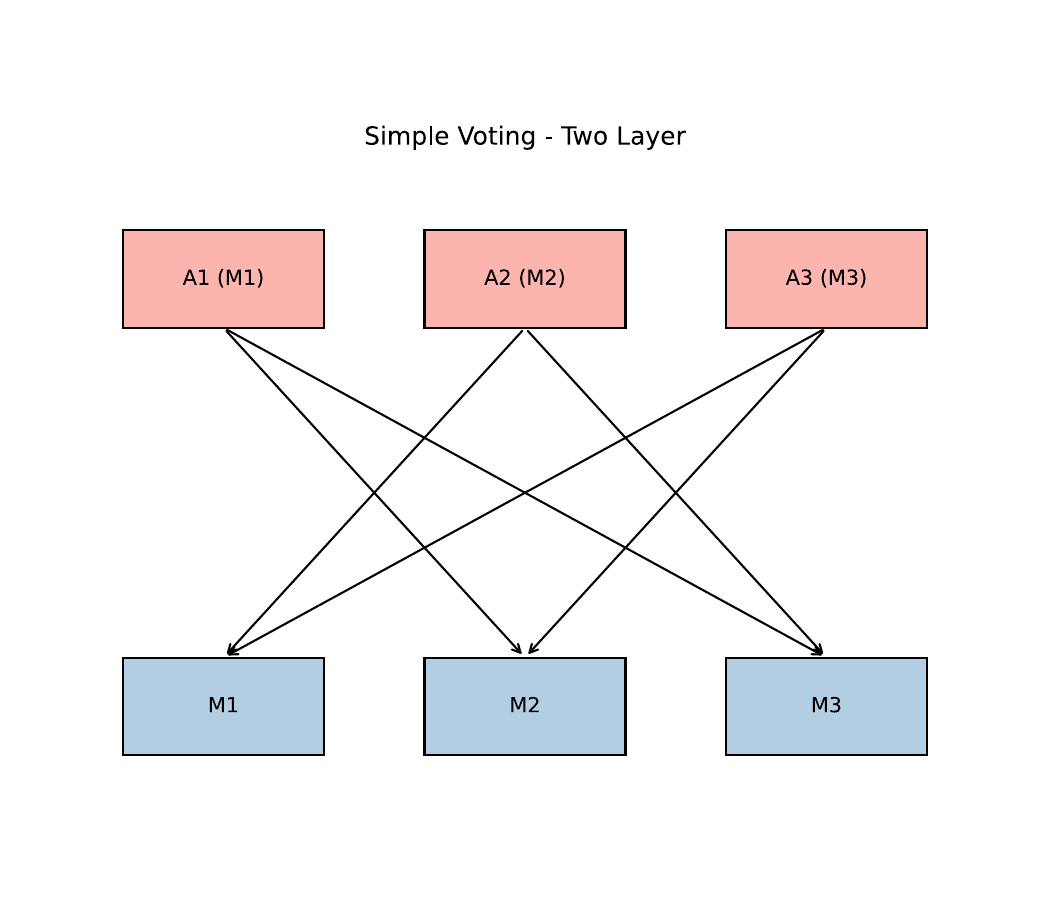}
    \caption{Simple voting-based gossip consensus flow.}
    \label{fig:simple-voting}
\end{figure}

\begin{lstlisting}[language=Python]
for model in MODELS:
    answer[model] = get_answer_and_thought_process(model, Q)

for model in MODELS:
    final_answer[model] = get_final_answer(model, answers, thoughts)

counter = {}
ANSWER = None
for model in MODELS:
    counter[final_answer[model]] += 1
    if counter[final_answer[model]] > count:
        ANSWER = final_answer[model]
\end{lstlisting}

\subsection{Voting Algorithm with a Judge}
\begin{itemize}
    \item One model is randomly chosen to act as the judge.
    \item All others submit their answers and thought processes to the judge.
    \item The judge selects the final answer from peer submissions.
    \item Judges are rotated across rounds to mitigate long-term bias.
\end{itemize}

\begin{figure}[ht]
    \centering
    \includegraphics[width=0.45\textwidth]{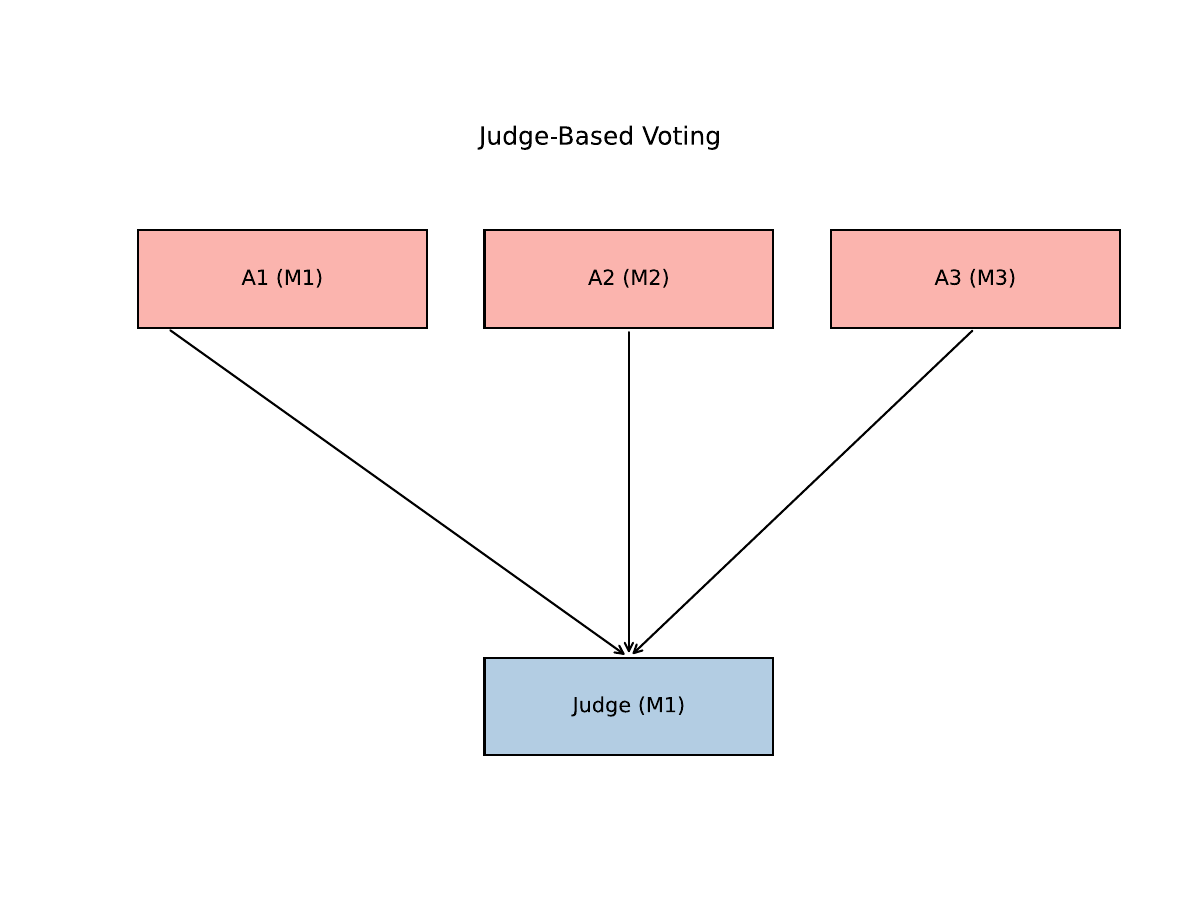}
    \caption{Consensus with a rotating judge.}
    \label{fig:judge-voting}
\end{figure}

\begin{lstlisting}[language=Python]
JUDGE = random.choice(MODELS)

for model in MODELS:
    if model != JUDGE:
        answer[model] = get_answer_and_thought_process(model, Q)

final_answer = get_answer(JUDGE, answers, thoughts)
\end{lstlisting}

\subsection{Multi-layer Consensus (Hierarchical)}
\begin{itemize}
    \item Models are grouped into sets that independently reach consensus.
    \item Each group selects a leader whose answer represents the group.
    \item Leaders then run a second-layer consensus round.
    \item Suitable for scaling to large LLM clusters and optimizing for context window limitations.
\end{itemize}

\begin{figure}[ht]
    \centering
    \includegraphics[width=0.45\textwidth]{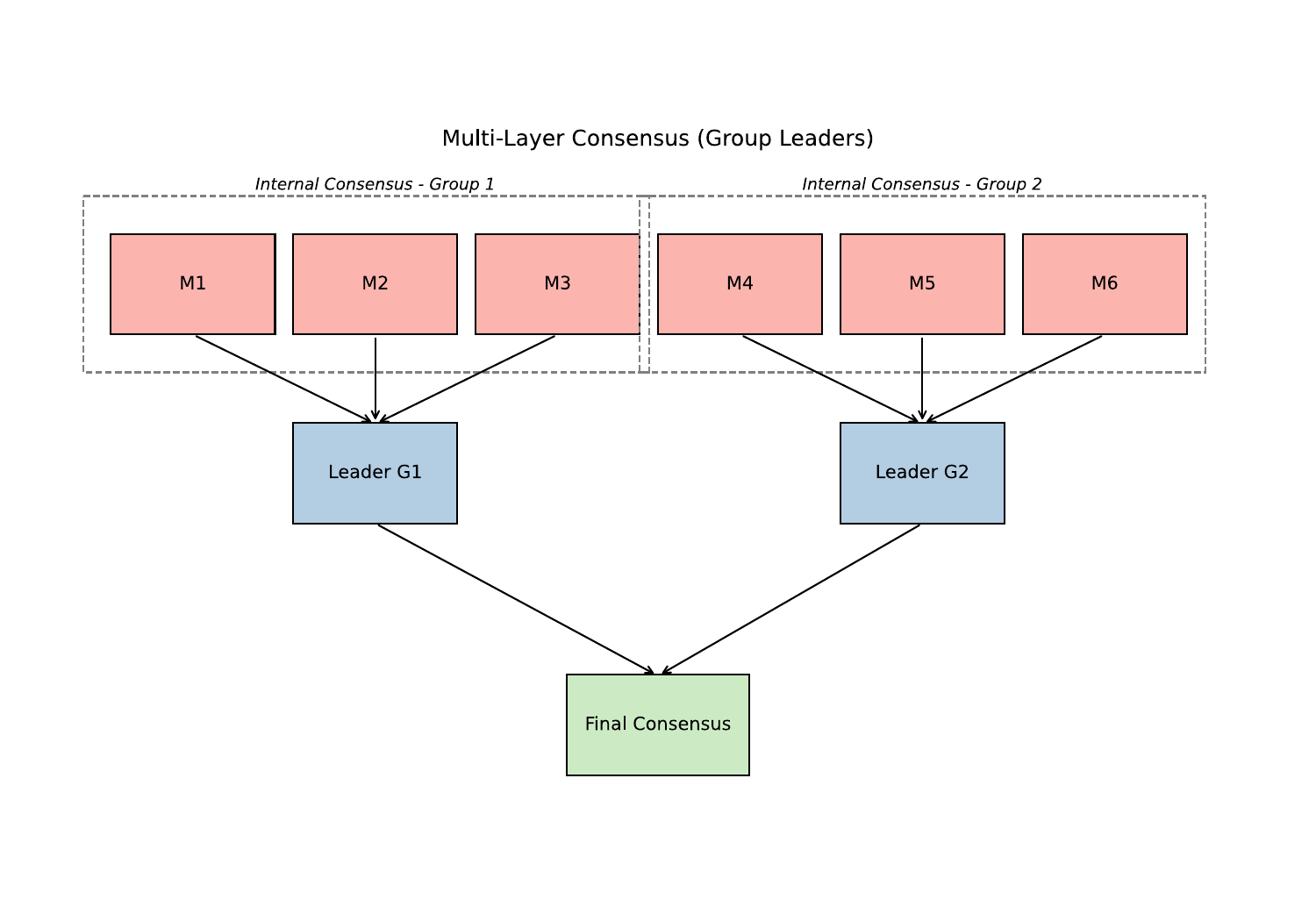}
    \caption{Multi-layer hierarchical consensus.}
    \label{fig:multi-layer}
\end{figure}

\begin{lstlisting}[language=Python]
GROUPS = { [m1, m2, m3], [m4, m5, m6] }
for group in GROUPS:
    for model in group:
        answer[group][model] = get_answer_and_thought_process(model, Q)
    leader = elect_leader(group, answer[group])
    leaders.append(leader)

ans = []
for leader in leaders:
    ans[leader] = get_answer_and_thought_process(leader, Q)

ANSWER = max(ans)
\end{lstlisting}

\section{Findings and Observations}

We tested our gossip-based consensus idea on two sets of models: one with stronger, newer models and one with low-end, lighter models. The goal was to see if letting them ``talk'' and then vote would beat their individual performance. All evaluations were run on a sample of \texttt{MMLU} benchmark \cite{mmlu2021} dataset.

\subsection{High-End Models}
The first group included \texttt{o4-mini}, \texttt{gemini-2.5-pro}, \texttt{grok-4-0709}, and \texttt{deepseek-reasoner}. On their own, these models scored between 82.6\% and 89.4\%. The best was \texttt{gemini-2.5-pro} at 89.4\%. When combined through majority voting, the accuracy jumped to 93.3\%.  

That is a relative improvement of about \textbf{+4.3 percentage points} over the best single model and \textbf{+10.7\% fewer errors}. Even when the models are already strong, consensus consistently squeezes out a little more performance.

\begin{figure}[ht]
    \centering
    \includegraphics[width=0.48\textwidth]{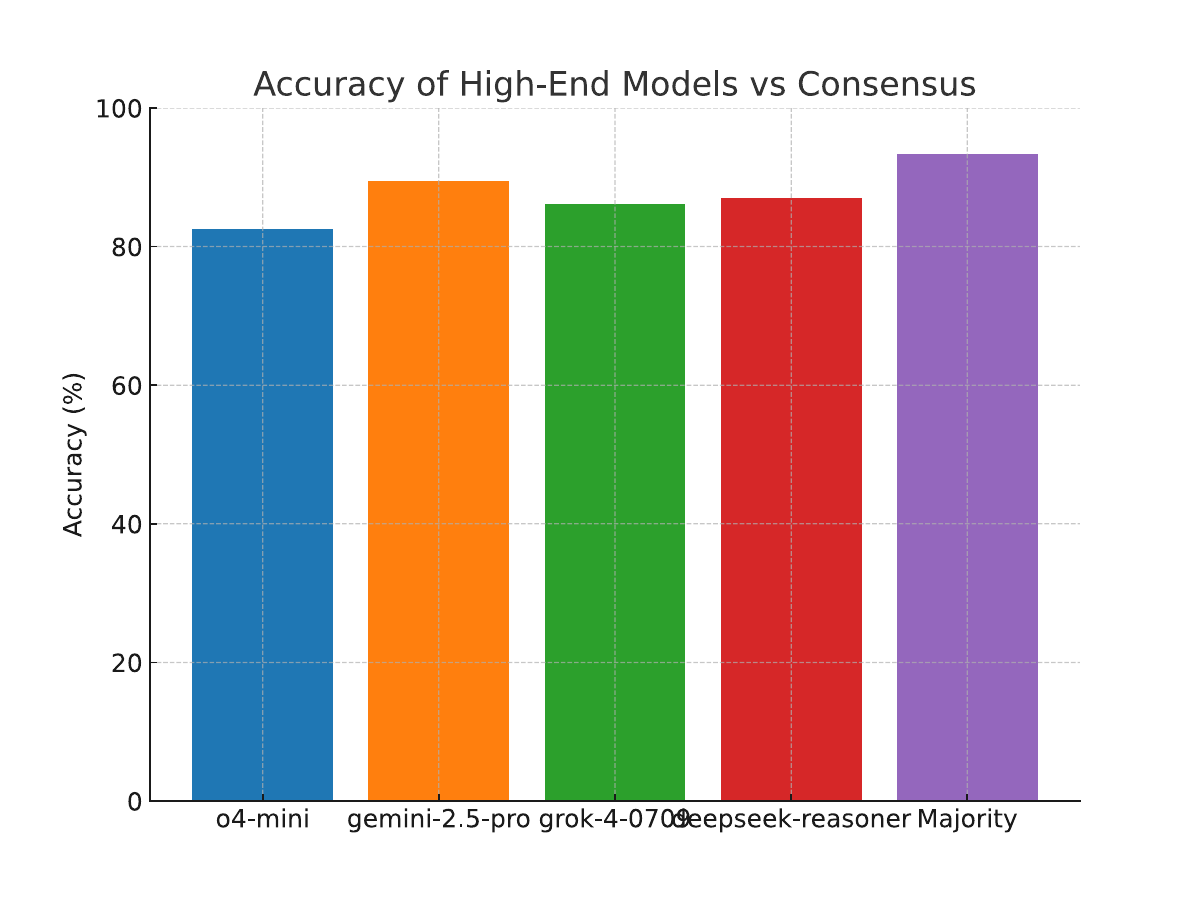}
    \caption{High-end models compared to their consensus.}
    \label{fig:high-end-results}
\end{figure}

\subsection{Low-End Models}
The second group had \texttt{o3-mini}, \texttt{gemini-1.5-flash}, \texttt{grok-3}, and \texttt{deepseek-chat}. These were weaker overall, scoring between 62.2\% and 77.3\%. The best single model was \texttt{grok-3} at 77.3\%. But when we ran consensus, accuracy went up to 84.2\%.  

That means a relative improvement of about \textbf{+6.9 percentage points} over the best single model and \textbf{+30.4\% fewer errors}. The effect here is much bigger compared to the stronger models, showing that consensus provides the largest benefit when the models are less reliable.  

Importantly, the total cost of running this entire low-end experiment was roughly \textbf{half the price of a single run with only \texttt{gemini-2.5-pro}}. This highlights a key point: consensus not only improves accuracy, it can do so while being \textit{cost-efficient}. With the right mix of models, we can achieve near frontier-level performance without always paying frontier-level prices.

\begin{figure}[ht]
    \centering
    \includegraphics[width=0.48\textwidth]{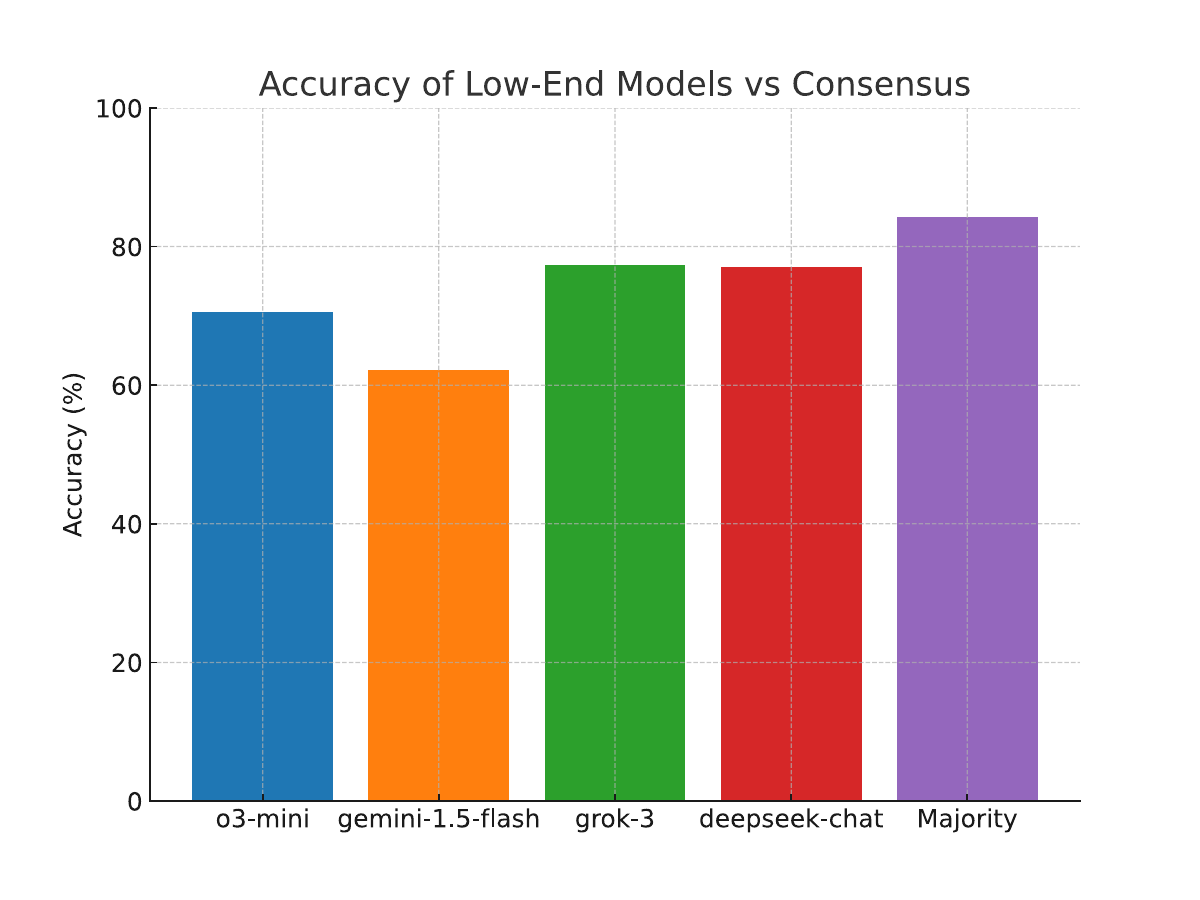}
    \caption{Low-end models compared to their consensus.}
    \label{fig:cheap-results}
\end{figure}

\subsection{Key Takeaways}
\begin{itemize}
    \item Consensus always beats the single best model in the group.
    \item The boost is modest for high-end models (+4.3 points, 10.7\% fewer errors), but large for low-end ones (+6.9 points, 30.4\% fewer errors).
    \item Consensus can be more cost-effective: the low-end set cost \textbf{50\% less} than a single \texttt{gemini-2.5-pro} run.
    \item Using a diversity of models is often better than relying on just one; each model brings different strengths.
    \item An automatic ``model chooser'' system could make this practical, routing questions to the most cost-efficient mix of models for consensus.
    \item Having multiple models balance each other also reduces mistakes and biases.
    \item The way the models agree feels closer to how humans reason together in groups.
\end{itemize}

In short, when models ``gossip'' and agree, they perform better than when they work alone. For weaker models, this makes a significant difference, and for stronger ones, it still adds reliability but also considerable cost and latency. Therefore, a smart way would be to combine diverse small models, which would increase accuracy and also decrease the overall cost, and hence, an affordable solution.

\subsection{Future Work}
While majority voting already shows strong improvements, there are several promising directions to extend this framework:

\begin{itemize}
    \item \textbf{Confidence-based consensus:} Weighted voting, where each model has given more influence, may seem attractive. However, it risks introducing long-term bias if a single model repeatedly dominates. A better approach is to make weights \textit{dynamic}, rewarding models when their answers align with ground truth and penalizing them when they are wrong. Over time, this would act like a training-like signal, producing a confidence-based consensus that balances fairness with accuracy.
    
    \item \textbf{Automatic model chooser:} Different models excel in different domains. For example, one model may be stronger in reasoning while another is better at factual recall. An automatic router could classify the incoming query and direct it to the most suitable set of models for consensus. This would maximize both accuracy and cost-efficiency by avoiding unnecessary use of frontier models.
    
    \item \textbf{Scaling to larger ensembles:} Our current experiments used four-model groups, but the gossip protocol naturally supports larger clusters. Future work could explore hierarchical or multi-layer gossip across dozens of heterogeneous models, allowing consensus to form at scale without exceeding context limits.
    
    \item \textbf{Human-AI hybrid consensus:} Since our framework mimics group deliberation, an intriguing direction is to include human reasoning as another "node" in the gossip. If ever the consensus is not found, a human can intervene and takeover the consensus. This could become critical for decision in fields such as healthcare, law, or governance, where human oversight is critical.
\end{itemize}

In short, consensus is not just a mechanism for combining models, but a foundation for building \textbf{adaptive, cost-aware, and trustworthy} multi-agent reasoning systems.

\end{document}